# Superconductivity of Tellurium Polyhydride with $T_c$ above 90K

Jinfu Zhu[1,2], Guiqi Liu[1,2], Yuanhao Su[1,2], Hongyu Liu[1,2], Sijia Zhang[1], Panpan Kong[1,2], Qingqing Liu[1], Jianfa Zhao[1], Shaomin Feng[1,2], Jun Zhang[1], Haoyu Zheng[1], Jing Song[1], Luhong Wang[3], Fuyang Liu[4], Haozhe Liu[4], M. Bykov[5], Xiancheng Wang[1,2] & Changqing Jin[1,2]

[1] *Beijing National Laboratory for Condensed Matter Physics, Institute of Physics, Chinese Academy of Sciences, Beijing 100190, China*

[2] *School of Physics, University of Chinese Academy of Sciences, Beijing 100190, China*

[3] *Shanghai Key Laboratory of Material Frontiers Research in Extreme Environments (MFree), Shanghai Advanced Research in Physical Sciences (SHARPS), Shanghai 201203, China*

[4] C*enter for High Pressure Science & Technology Advanced Research (HPSTAR), Beijing, 100094, China*

*5 Department of Chemistry, Johann Wolfgang Goethe University, Germany*

**Abstract**

The experimental realization of high-temperature superconductivity (SC) in compressed $SH_3$ has greatly stimulated the exploration of novel polyhydride superconductors. Nevertheless, aside from the landmark discovery of $SH_3$, experimental investigations of other chalcogen polyhydride superconductors remain unreported to date. Here, we report the first experimental observation of SC in tellurium (Te) polyhydride. The compound was synthesized at high pressure and high temperature conditions using a diamond anvil cell combined with a laser heating system. Subsequent in situ transport measurements at high pressures, performed as a function of temperature and applied magnetic field, revealed a superconducting transition with a critical temperature $T_c$ about 91 K at 263 GPa. The superconducting phase is assigned to $TeH_4$ with characterized face shared $TeH_{12}$ cage forming quasi molecular $H_2$ units based on synchrotron x-ray diffraction experiments. Analysis of the SC behavior at magnetic fields yielded a Ginzburg Landau (GL) coherence length of approximately $\xi \sim 54$ Å. Tellurium polyhydride thus becomes another chalcogen polyhydride superconductor in addition to the landmark discovery of sulfur hydride.

## Introduction

The search for new high-temperature superconductors in polyhydrides has intensified recently, motivated by the "pre-compression" effect that could efficiently lower hydrogen metallization pressures to experimentally achievable levels [1-11]. Following the theoretical prediction of high-temperature superconductivity (SC) in sulfur hydrides [4, 5], experimental confirmation soon followed when A. P. Drozdov et al. discovered that $SH_3$ exhibits a superconducting transition temperature $T_c$ of up to 203 K at 155 GPa [12]. This landmark discovery of high-$T_c$ SC in $SH_3$ stimulated extensive experimental efforts to search for other binary hydride superconductors [13-27]. Notable successes include the discoveries of SC at megabar pressures in $LaH_{10}$ with $T_c$ = 250-260 K [13, 15], $YH_9$ with $T_c$ = 243 K [16] and $CaH_6$ with $T_c$ of ~210 K [17, 18]. Beyond these rare-earth and alkali-earth polyhydrides with $T_c$ exceeding 200 K, SC has also been experimentally observed in other metal polyhydrides with moderate transition temperatures [19-25], such as $ThH_{10}$ with $T_c$ = 161 K at 175 GPa [20]. Among heavy rare-earth elements with fully filled $f$-shells, lutetium hydride ($Lu_4H_{23}$) was reported to exhibit SC with a $T_c$ of 71 K at 218 GPa [21]. In contrast, for rare-earth elements with partially filled $f$-shells, SC in their hydrides is often significantly suppressed by local moments arising from the $f$-electrons [28, 29]. Hydrides of Group IVB and VB metals have also been found to be superconducting at megabar pressures, with $ZrH_6$ showing a $T_c$ of 71 K [22], $HfH_{14}$ 83 K [23], $TaH_3$ 30 K [25], and $NbH_3$ 42 K [27]. For elements on the right side of the periodic table, besides the well-known $SH_3$ ($T_c$ =203 K), superconductivity has been reported in $SnH_n$ ($T_c$ ~70 K) [24], $SbH_4$ ($T_c$ ~ 116 K) [26], $BiH_4$ ($T_c$ ~ 91 K) [30] and $BiH_2$ ($T_c$ ~ 62 K) [31]. In addition to binary hydrides, ternary hydrides have also been experimentally realized. For instance, $LaBeH_8$ ($T_c$ ~ 110 K) [32] and $LaBH_8$ ($T_c$ ~ 106 K) [33] have been synthesized and found to be SC at pressures below 100 GPa.

Although the experimental discovery of SC in $SH_3$ stands as a landmark in the field of hydride superconductors, SC in hydrides of the other Group-VIA elements, selenium and tellurium, has not been experimentally reported prior to this work. Here, we report on another chalcogen hydride superconductor: tellurium polyhydride, in which SC with a $T_c$ of up to 91 K was experimentally observed.

## Experimental details

Tellurium polyhydride samples were synthesized at high pressure and high temperature using a laser-heated diamond anvil cell (DAC). For megabar experiments, we used diamond anvils with a 50 μm culet beveled to 300 μm. A T301 stainless steel gasket was pre-indented to ~10 μm, drilled with a 300 μm hole, filled with densely pressed $Al_2O_3$, and then re-drilled to create a 40 μm sample chamber. Ammonia borane (AB) served as both the hydrogen source and pressure medium. Platinum inner electrodes were deposited on the anvil culet, onto which a tellurium foil ($20 \times 20 \times 1$ μ$m^3$) was placed. Pressure was determined from the diamond Raman edge shift. Experimental details follow the ATHENA procedure described in Ref. [34].

In situ laser heating of the high-pressure sample was performed using a 1064 nm YAG laser with a beam diameter of ~5 μm. The sample was heated at ~2000 K for several seconds to decompose ammonia borane (AB), releasing hydrogen to react with tellurium and form tellurium polyhydride. Heating temperatures were determined from blackbody radiation fitting. The synthesis pressure was maintained for subsequent electrical transport measurements, which were conducted in a MagLab system from 300 K to 1.5 K and under magnetic fields up to 5 T. Resistance was measured using the Van der Pauw method [35-37] with an applied current of 0.1 mA.

In-situ high-pressure X-ray diffraction (XRD) experiments were conducted at Shanghai Synchrotron Radiation Facility with a radiation wavelength of 0.6199 Å. Symmetric diamond anvil cells (DACs) with rhenium gaskets were used, with a sample chamber diameter of ~25 μm. Pressure was determined from the diamond Raman edge shift. Two-dimensional XRD images were reduced to one-dimensional patterns using the Dioptas software package.

## Results and discussions

Figure 1 shows the temperature-dependent resistance of the tellurium polyhydride sample (Cell #1), measured at 263 GPa under different magnetic fields, the same pressure at which it was synthesized. The resistance displays two distinct drops upon cooling, suggestive of two phase-transition events. The inset presents an enlarged view of the $R(T)$ curve, highlighting the first resistance drop at 91 K. The resistance approaches zero at low temperatures, and these transitions are substantially suppressed by applied magnetic field. These observations exclude the possibility that the resistance drops arise from structural or magnetic phase transitions, implying superconductivity (SC) with an onset critical temperature $T_c$ ~ 91 K. The two-step superconducting transitions most likely originate from tellurium polyhydride phases with variable hydrogen content, a phenomenon commonly reported for other polyhydride superconductors [13, 17, 18].

For a second tellurium polyhydride sample (Cell #2), resistance measurements were conducted at its synthesis pressure of 229 GPa (Fig. 2). Zero resistance is achieved below 15 K for this sample, as clearly demonstrated in the lower inset of Fig. 2. The onset $T_c$, shown in the upper inset, is approximately 82 K, slightly lower than that of Cell #1. It appears that $T_c$ rises with increasing pressure for tellurium polyhydride samples synthesized above 200 GPa. Nevertheless, pressure-cycling experiments to probe the pressure dependence of SC could not be performed, as we routinely implemented for other polyhydride systems [17, 25]. Synthesizing tellurium polyhydride proved exceptionally challenging owing to severe difficulties in laser-heating the sample. The high laser power required for synthesis frequently damaged the diamond anvils, precluding such pressure-cycling measurements.

The magnetic-field dependence of the superconducting transition was investigated. For the Cell #1 sample, the temperature dependence of the upper critical magnetic field $\mu_0 H_{c2}(T)$ is plotted in Fig. 3. The onset $T_c$ decreases linearly from 91 K to ~69 K as the magnetic field rises to 4.8 T. A linear fit to the data, presented in the inset of Fig. 3, gives a slope of $|dH_c/dT|$= 0.22 T/K near $T_c$. According to the Werthamer-Helfand-Hohenberg (WHH) theory, the zero-temperature upper critical field governed by the orbital depairing mechanism in the dirty limit, $\mu_0 H_{c2}^{\mathrm{Orb}}(0)$, can be estimated using the formula $\mu_0 H_{c2}(T) = -0.69\times[dH_{c2}/dT|_{Tc}]\times T_c$. Using the slope of −0.22 T/K and the onset $T_c$=91 K, we obtain $\mu_0 H_{c2}^{\mathrm{Orb}}(0)$ ~13.8 T. The zero-temperature upper critical field was also estimated using the Ginzburg-Landau (GL) theory, $\mu_0 H_{c2}(T) = \mu_0 H_{c2}^{\mathrm{GL}}(0)(1-(T/T_c)^2)$. Fitting this equation to the $\mu_0 H_{c2}(0)$ data yields $\mu_0 H_{c2}^{\mathrm{GL}}(0)$ ~11.2 T, which is comparable to the orbital-limited value $\mu_0 H_{c2}^{\mathrm{Orb}}(0)$. For comparison, the Pauli-limiting upper critical field due to the spin depairing mechanism (Zeeman effect) for a weak-coupling superconductor is given by $\mu_0 H_{c2}^{\mathrm{P}}(0) = 1.86\times T_c$. Using the same onset $T_c$=91 K, this yields $\mu_0 H_{c2}^{\mathrm{P}}(0)$ ~ 169 T. The significantly smaller value of $\mu_0 H_{c2}^{\mathrm{Orb}}(0)$ compared to $\mu_0 H_{c2}^{\mathrm{P}}(0)$ indicates that Cooper pair breaking is predominantly governed by the orbital depairing mechanism. From $\mu_0 H_{c2}^{\mathrm{GL}}(0)= \Phi_0/2\pi\xi^2$ with flux quantum $\Phi_0$= $2.067\times10^{-15}$ Web, the GL coherence length is estimated to be $\xi$ ~

54 Å.

Figure 4(a) shows the resistivity of sample Cell #2 measured under different magnetic fields. The inset of Fig. 4(a) illustrates that the onset $T_c$ decreases from 82 K to 67 K as the magnetic field rises to 5 T. The zero-resistance state present at zero field is also suppressed upon application of a magnetic field. This behavior points to weak superconducting links between crystalline grains within the synthesized sample. At high magnetic fields, field penetration into the sample disrupts these weak links and quenches zero-resistance behavior, a hallmark widely reported for granular superconductors. Figure 4(b) plots the upper critical magnetic field as a function of temperature for sample Cell #2. The data can be well described by a linear fit, yielding a slope of -0.31 T/K. Although Cell #2 exhibits a lower $T_c$ than Cell #1 (measured at higher pressure), its absolute slope $|dH_c/dT|$ is higher. This suggests that the SC of the higher-pressure Cell #1 is more susceptible to magnetic-field suppression. This phenomenon originates intrinsically from high-pressure effects. Applied compression broadens electronic bandwidths, suppressing both the Fermi-surface density of states $N(E_F)$ and the normal-state resistivity $\rho_n$. Within the dirty limit of the WHH theory, the slope near $T_c$ scales proportionally to the product $N(E_F)\cdot\rho_n$. Consequently, a reduced value of this product under higher pressure gives rise to a smaller slope magnitude.

Superconductivity in tellurium hydrides has been investigated theoretically [38]. Three metallic stoichiometries of $TeH_4$, $Te_2H_5$, and TeH were predicted to be stable up to 250 GPa, with estimated $T_c$ of ~99 K, ~58 K and 19 K at 200 GPa, respectively. The predicted $T_c$ for the hexagonal $P6/mmm$ phase of $TeH_4$ is comparable to our experimentally observed value, suggesting that the first superconducting transition observed in our samples likely originates from $TeH_4$. To experimentally identify the superconducting phase, we conducted preliminary in-situ high-pressure X-ray diffraction experiments at 209 GPa for the sample Cell #3. The diffraction pattern presented in Fig. 5(a) confirms the presence of the hexagonal $TeH_4$ phase, with refined lattice constants $a$ = 2.9876 Å and $c$ = 2.6488 Å that are in good agreement with theoretical predictions [38]. Based on structural predictions, the crystal structure of the hexagonal $P6/mmm$ phase of $TeH_4$ is illustrated in Fig. 5(b). The structure consists of face-sharing $TeH_{12}$ cages, and the nearest neighbor H atoms are covalently bonded between adjacent layers, forming quasi-molecular $H_2$ units. The bonding between Te and H is suggested to be primarily ionic in character [38], attributable to the larger atomic core and weaker electronegativity of Te compared to S and Se.

## Conclusion

In summary, tellurium polyhydride has been successfully synthesized at megabar pressures and found to exhibit SC with a $T_c$ up to 91 K at 263 GPa. The upper critical field $\mu_0 H_{c2}(0)$ was estimated to be ~11 T. In-situ high-pressure X-ray diffraction experiments reveal the presence of a hexagonal $TeH_4$ phase, which is likely responsible for the observed SC. Tellurium polyhydride thus becomes the second chalcogen polyhydride superconductor, following the landmark discovery of sulfur hydride.

**Figure Captions:**

Figure 1. $R(T)$ curves of the tellurium polyhydride sample (Cell #1) measured at 263 GPa under different magnetic fields. The inset shows an enlarged view highlighting the onset $T_c$.

Figure 2. Temperature dependence of resistance for tellurium polyhydride sample (Cell #2) measured at 229 GPa. The upper and lower insets are the enlarged $R(T)$ curves to show the onset $T_c$ and zero resistance, respectively.

Figure 3. The temperature dependence of upper critical magnetic field $\mu_0 H_{c2}(T)$. The red line denotes the GL fitting result. The inset displays the corresponding linear fit.

Figure 4. (a) $R(T)$ curves measured at different magnetic fields for tellurium polyhydride sample (Cell #2). The inset is the enlarged curves illustrating the onset $T_c$ at different fields. (b) Upper critical magnetic field $\mu_0 H_{c2}(T)$ as a function of temperature, together with the linear fit.

Figure 5. (a) X-ray diffraction pattern collected at 209 GPa for the tellurium polyhydride sample (Cell #3). (b) Schematic illustration of the crystal structure for the hexagonal P6/*mmm* phase of $TeH_4$ and the $TeH_{12}$ cage.

Figure 1

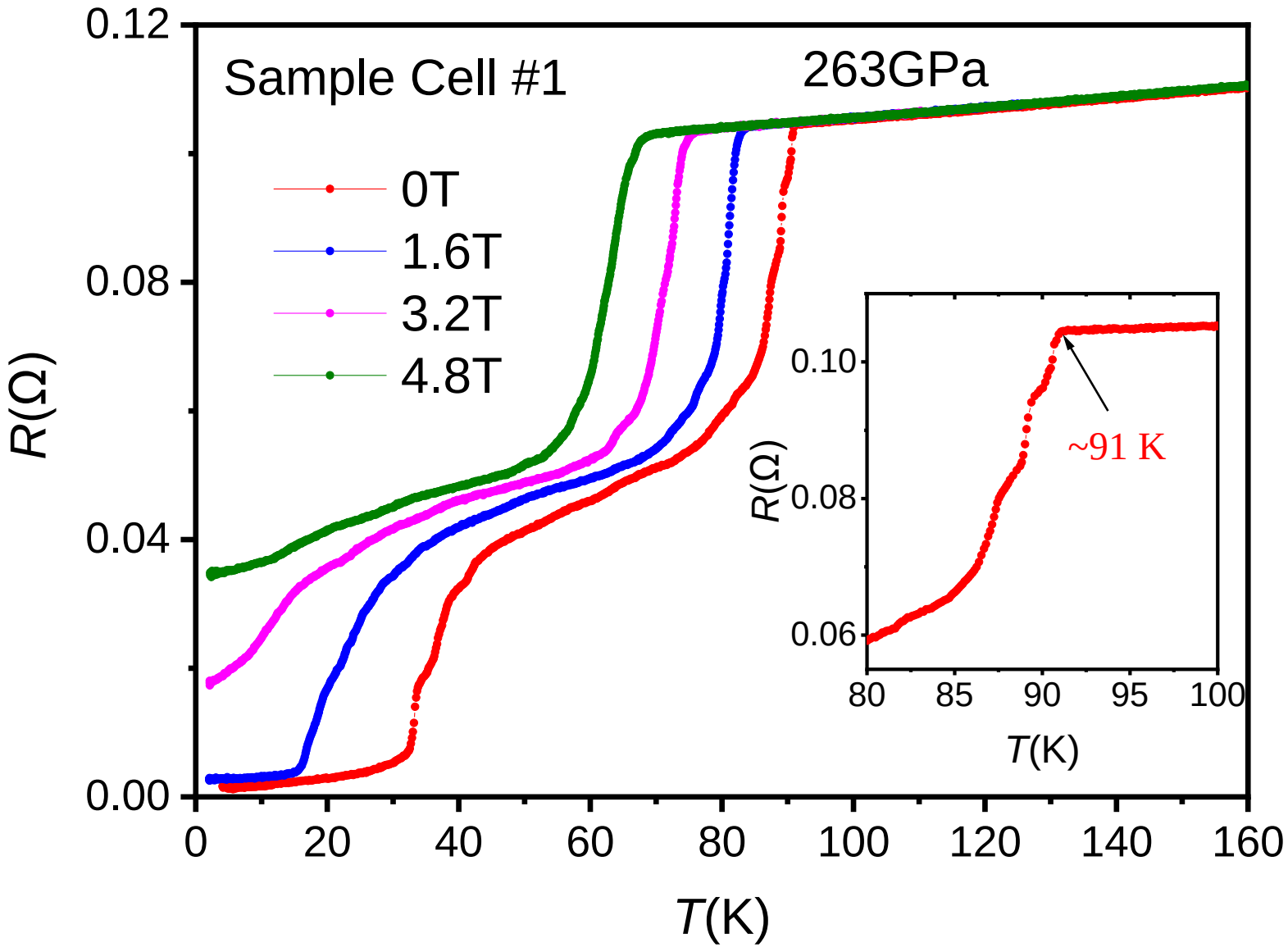

Figure 2

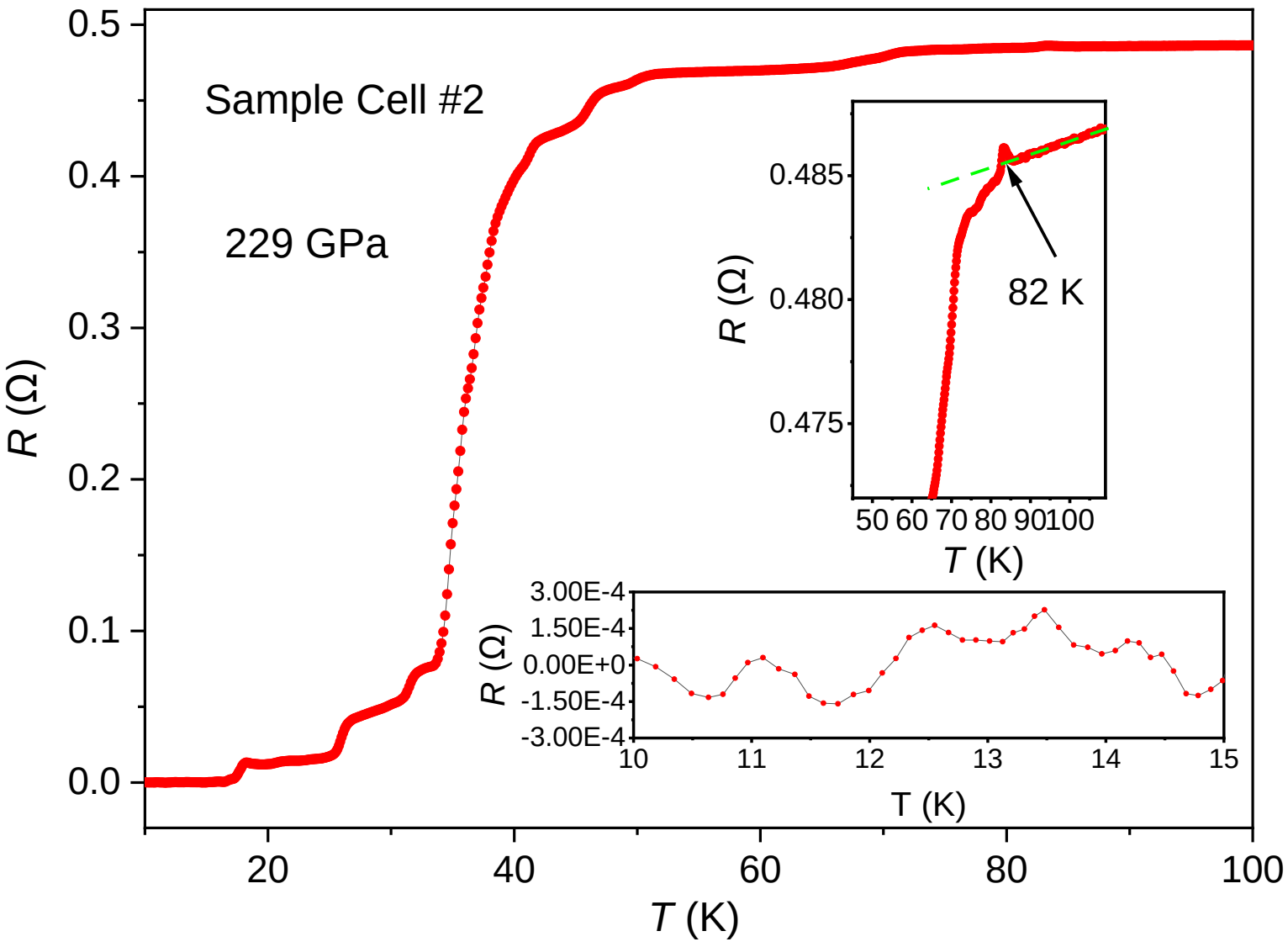

Sample Cell #2
229 GPa
R (Ω)
T (K)
0.0
0.1
0.2
0.3
0.4
0.5
20
40
60
80
100
0.475
0.480
0.485
50 60 70 80 90100
82 K
3.00E-4
1.50E-4
0.00E+0
-1.50E-4
-3.00E-4
10
11
12
13
14
15

Figure 3

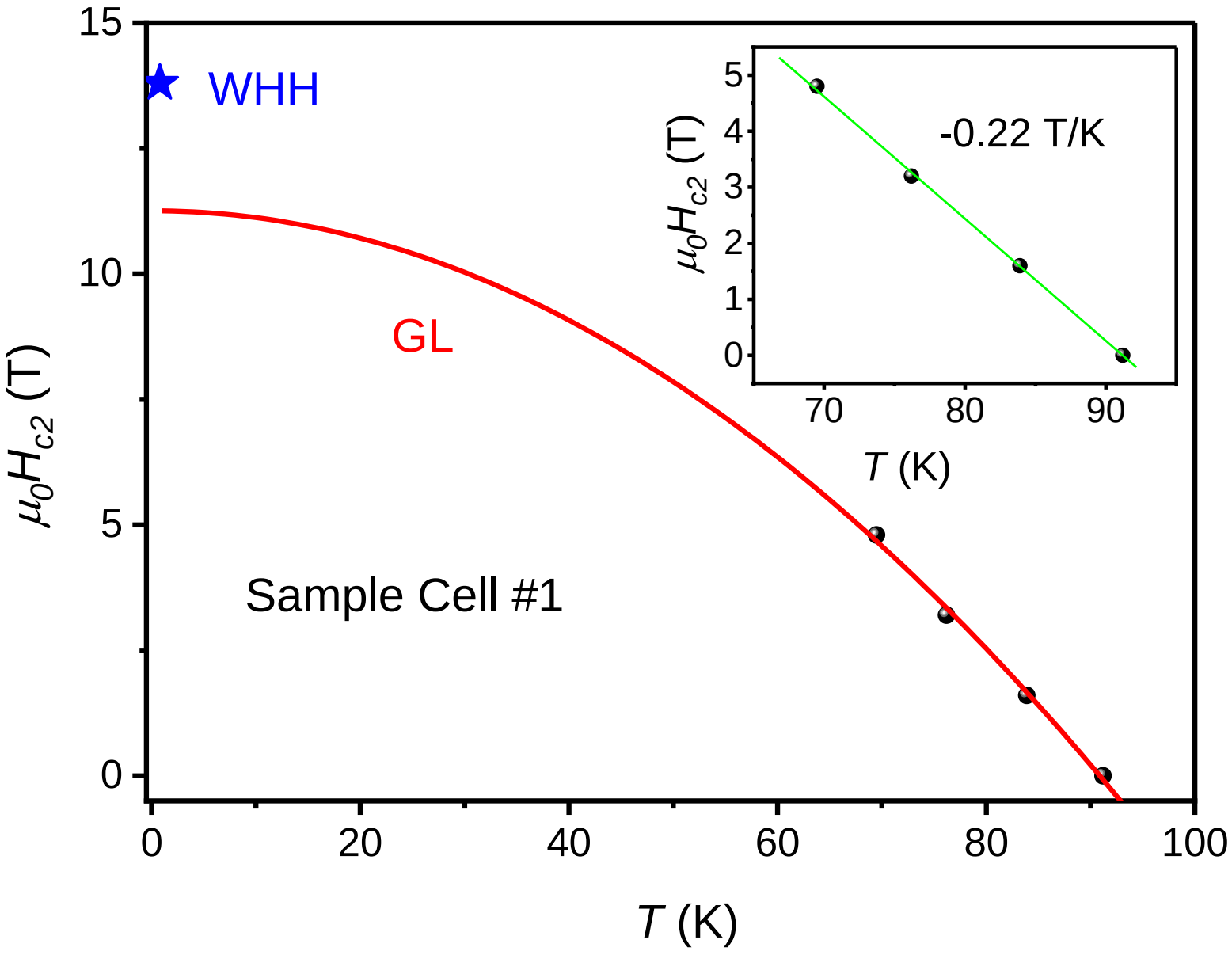

Figure 4

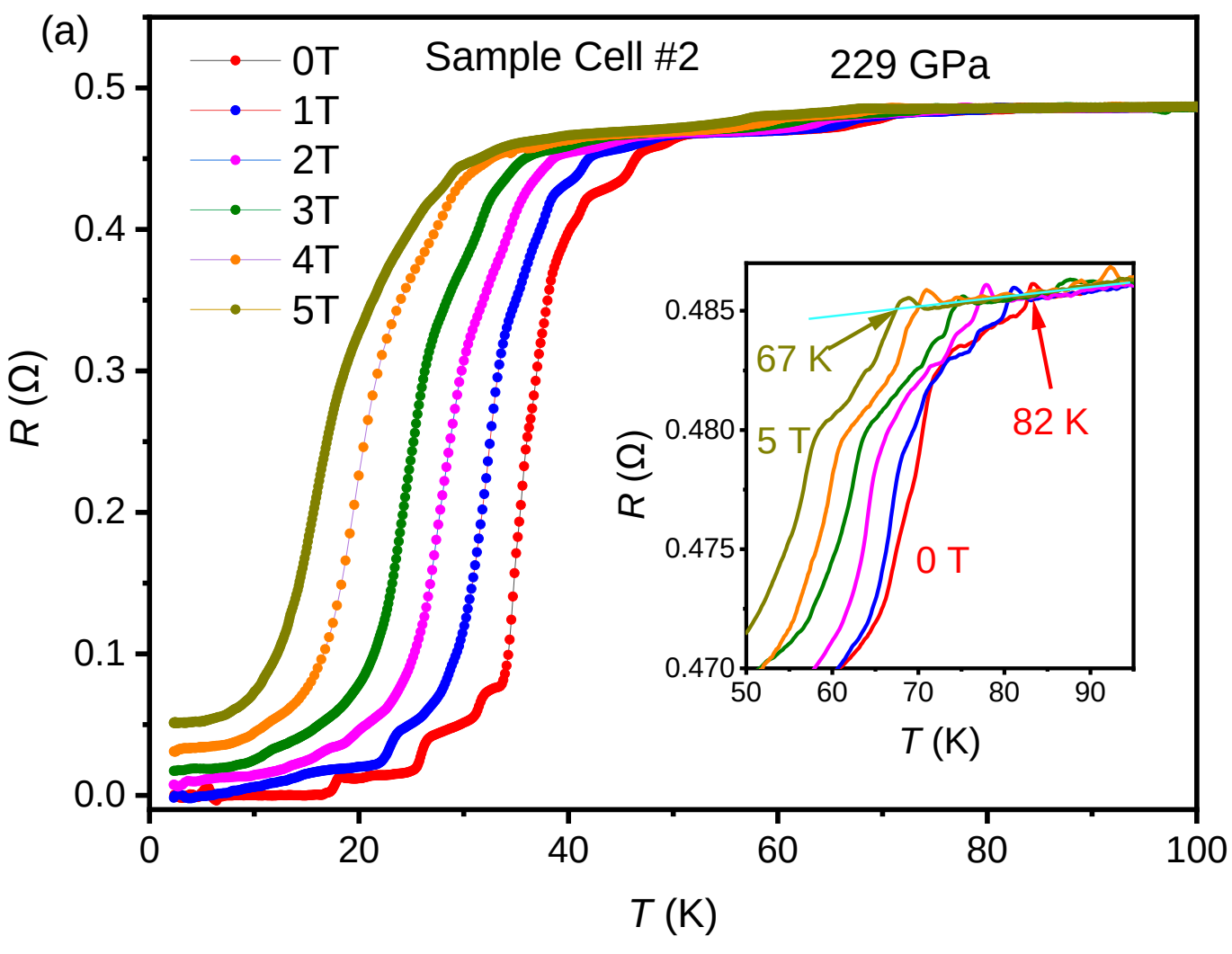
(a)
0T
1T
2T
3T
4T
5T
Sample Cell #2
229 GPa
R (Ω)
T (K)
67 K
5 T
82 K
0 T

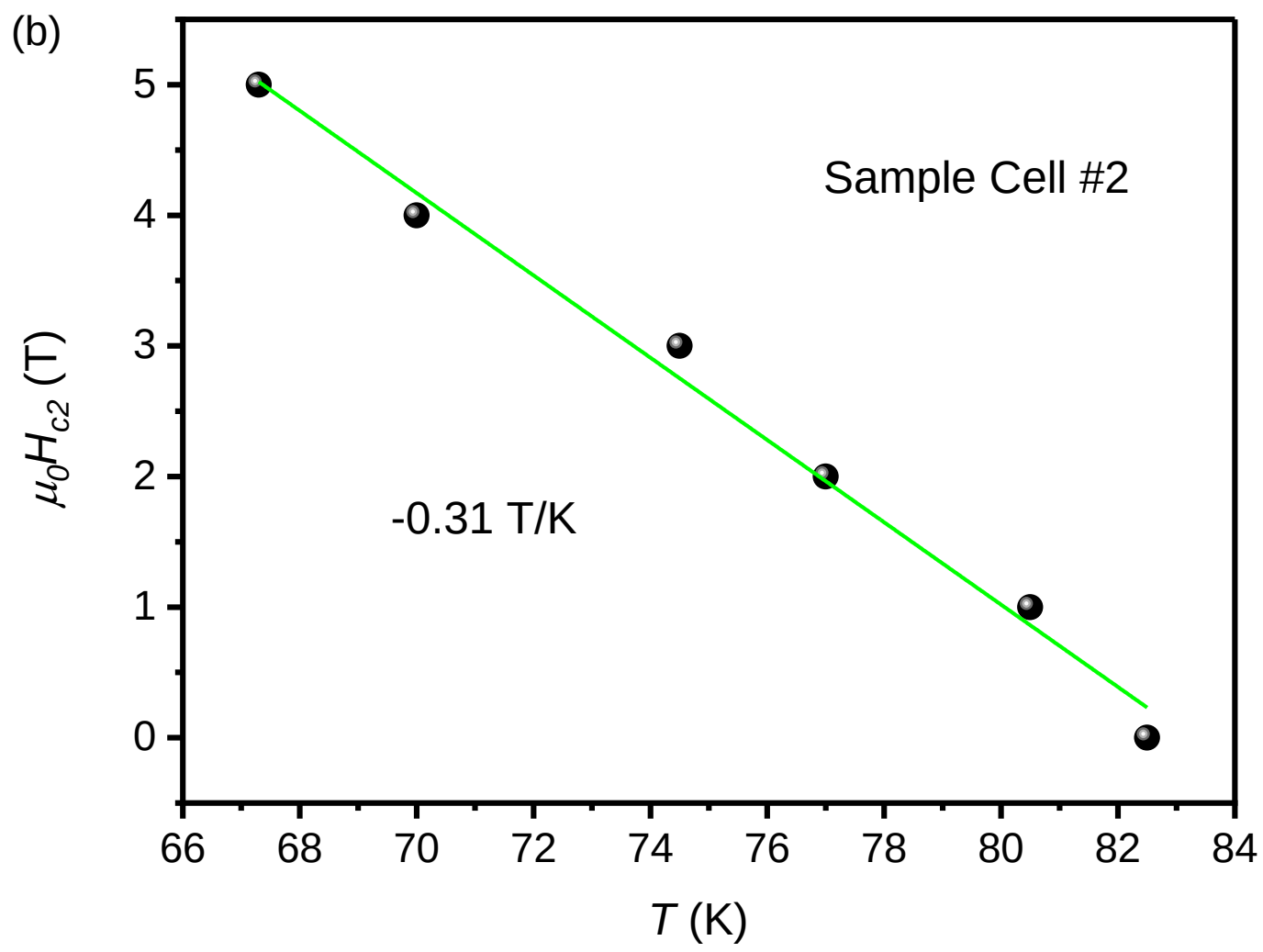
(b)
Sample Cell #2
-0.31 T/K
μ0Hc2 (T)
T (K)

Figure 5

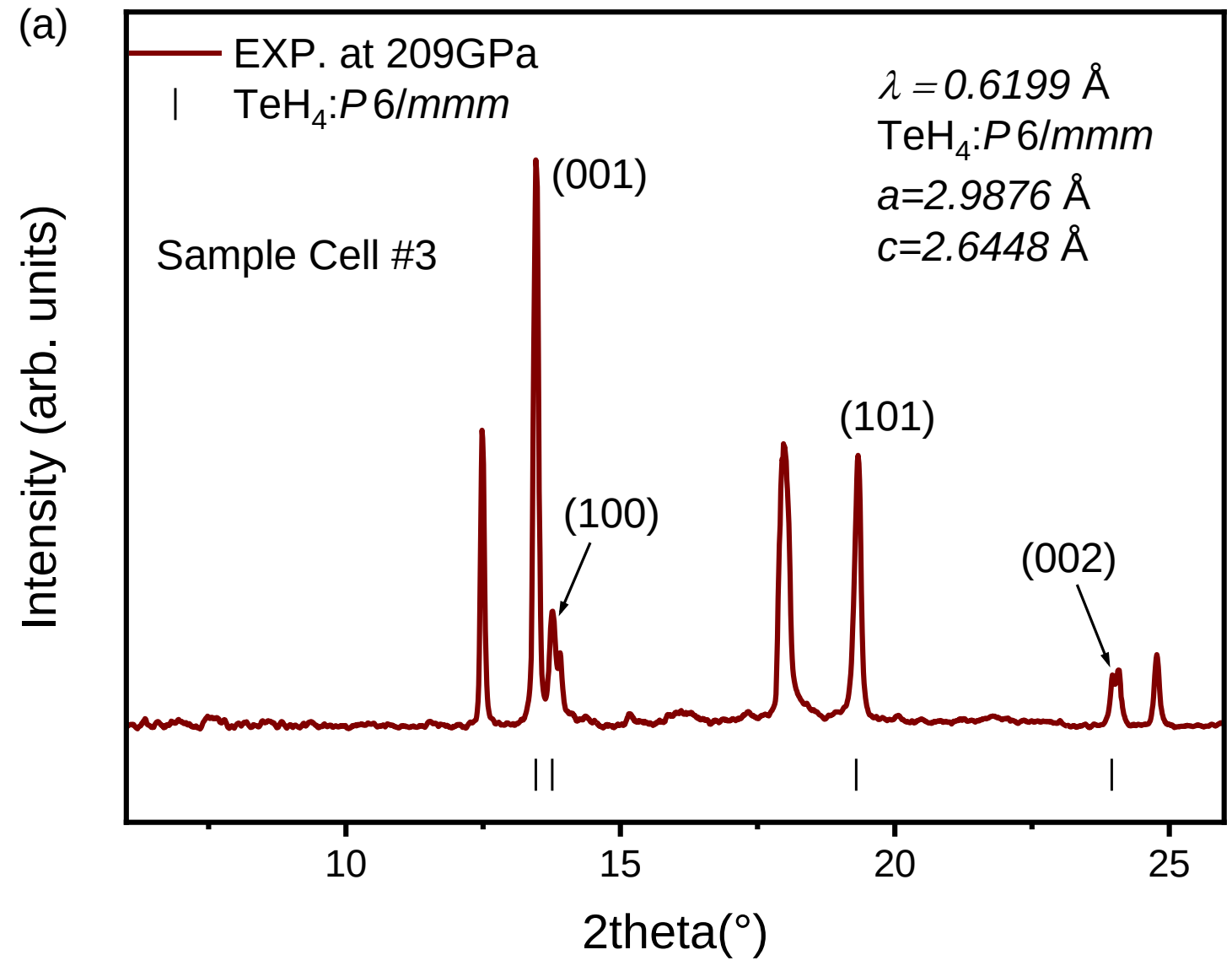

(a)
EXP. at 209GPa
TeH4:P6/mmm
λ = 0.6199 Å
TeH4:P6/mmm
a=2.9876 Å
c=2.6448 Å
Sample Cell #3
(001)
(100)
(101)
(002)
Intensity (arb. units)
10
15
20
25
2theta(°)

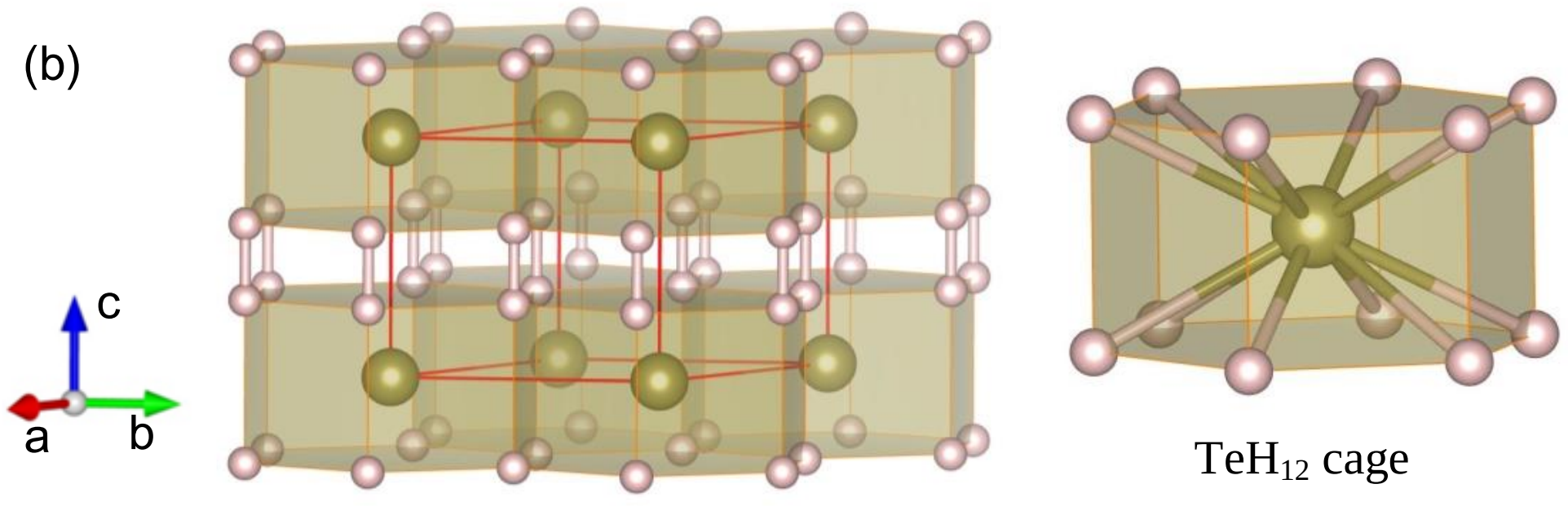

(b)
c
a
b
TeH12 cage